\documentclass{aa}
\usepackage{epsf,latexsym,here}
\usepackage{graphicx}

\begin{document}
\title{Studying the coincidence excess between EXPLORER and NAUTILUS 
       during 1998}
\author{D. Babusci \inst{1} 
   \and G. Cavallari \inst{2}
   \and G. Giordano \inst{1} 
   \and I. Modena \inst{3,}\inst{4}
   \and G. Modestino \inst{1}
   \and A. Moleti \inst{3,}\inst{4}
   \and G. Pizzella \inst{3,}\inst{1}
   }
\offprints{D. Babusci, 
\email{danilo.babusci@lnf.infn.it}}
\institute{INFN, Laboratori Nazionali di Frascati, via Enrico Fermi 40, 
           I-00044 Frascati (Roma), Italy\\
	\and CERN, CH-1211 Geneva 23, Switzerland\\ 
	\and Dipartimento di Fisica dell'Universit\`a ``Tor Vergata", 
	     via della Ricerca Scientifica 1, I-00133 Roma, Italy\\
	\and INFN, Sezione di Roma II, via della Ricerca Scientifica 1, 
	     I-00133 Roma, Italy}
\date{Received ..../ Accepted ....}
\abstract{The coincidences between EXPLORER and NAUTILUS during 
          1998 (Astone et al. 2001) are more deeply studied. It 
	    is found that the coincidence excess is greater in the 
	    ten-day period 7-17 September 1998 and it occurs at the 
	    sidereal hour 4, when the detectors axes are perpendicular 
	    to the Galactic Disc. The purpose of this paper is to bring 
	    our results with the GW detectors to the attention of 
	    scientists working in the  astrophysical field, and ask 
	    them whether are they aware of any special phenomenon 
	    occurring when EXPLORER and NAUTILUS showed a coincidence 
	    excess.
         }
\titlerunning{EXPLORER and NAUTILUS detectors}
\authorrunning{D. Babusci et al.}
\maketitle
%
%
%
\section{Introduction}
Since the initial claims by Weber (Weber 1969), which were not confirmed with successive 
experiments by other groups, the underlying idea has been to consider the gravitational wave 
(GW) emission as a phenomenon somewhat uniform in time. 

The search for short GW bursts within the IGEC collaboration (Allen et al. 2000, 
Astone et al. 2003) covering the period 1997-2000 produced upper limits for the GW flux 
over extended periods of time. The ROG collaboration has also presented the results obtained 
with the EXPLORER and NAUTILUS cryogenic bar detectors alone in the years 1998 (Astone et al. 2001) 
and 2001 (Astone et al. 2002). For 1998 the EXPLORER and NAUTILUS data show a small coincidence 
excess with a data selection favouring the Galactic Centre. An excess of events with respect 
to the expected background was found also in 2001, concentrated around sidereal hour four, 
when the two bars are oriented perpendicularly to the galactic plane, and therefore their 
sensitivity for galactic sources of GW is maximal. 

If the GW emission is a \emph{local} (in time as well in space) phenomenon, as, for instance, 
a supernova in a galaxy or a magnetar like that in December 2004, one should also investigate 
whether a deviation from the background occurs in relatively short periods of time.

A deviation from the background has, indeed, occurred in 1998 for the EXPLORER/NAUTILUS experiment, 
as we shall discuss in the following. This paper is especially addressed to astronomers, asking 
whether they may have observed peculiar phenomena at the same time.

\section {Experimental data}
During 1998 the resonant mass GW detectors NAUTILUS, installed at the INFN Frascati Laboratory, 
and EXPLORER, installed at CERN, operated from 2 June to 14 December for a common total measuring 
time of 94.5 days. Both detectors consist of an aluminium cylindrical bar having a mass of 2.3 
tons.The principle of operation of these detectors is based on the idea that the GW excites the 
first longitudinal mode of the bar, which is isolated from seismic and acoustic disturbances and 
is cooled to cryogenic temperatures to reduce the thermal noise. To measure the strain of the bar, 
a capacitive resonant transducer, tuned to the cited mode, is mounted on one bar face, followed 
by a very low noise superconducting amplifier.

The data are filtered with an adaptive filter matched to delta-like signals for the detection 
of short bursts (Astone et al. 1997). The variance of the filtered data is called 
\emph{effective temperature} and is indicated with $T_{\rm eff}$. In order to extract from the 
filtered data sequence $events$ to be analyzed we set a threshold at $E_{\rm thr} = 19.5\,T_{\rm eff}$. 
When the signal energy goes above the threshold, its time behaviour is considered until it falls back 
below the threshold for longer than a \emph{waiting time} of ten seconds\footnote{In this paper we 
study in more detail the results published in our previous paper (Astone et al. 2001), thus we 
maintain the same definition of \emph{event}. The events used for IGEC were obtained with a 
different threshold and a different waiting time.}. The maximum energy $E_s$ and its occurrence 
time define the \emph{event}.                                                           

Computation of the GW amplitude $h$ from the energy signal $E_s$ requires a model for the signal 
shape. A conventionally chosen shape is a short pulse lasting a time of $\tau_g$, resulting in 
the relationship
\begin{equation}
h \,=\, \frac{L}{v^2}\,\frac{1}{\tau_g}\,\sqrt{\frac{kE_s}{M}}, 
\label{he}
\end{equation}
where $v = 5400$ m/s is the sound velocity in aluminium, $L$ and $M$ the length and the mass of 
the bar and $\tau_g$ is conventionally assumed equal to 1 ms (for instance, for $E_s = 10$ mK 
we have $h = 7\,\times\,10^{-19}$ which requires, using the classical cross-section, a total 
conversion into GW of about $10^{-3}$ solar masses at the Galactic Centre). 

The main characteristics of EXPLORER and NAUTILUS in 1998 are reported in Table 1 of the paper 
Astone et al. 2001.

\begin{figure}
   \resizebox{\hsize}{!}{\includegraphics{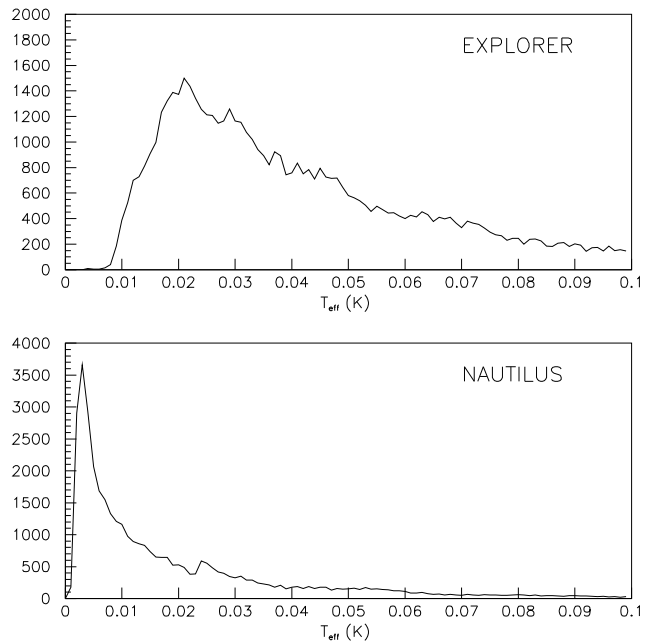}}
   \caption{Distribution of $T_{\rm eff}$ in kelvin units during 1998 
            (one value for each event). The average values of $T_{\rm eff}$ 
		are 41 mK for EXPLORER and 19 mK for NAUTILUS.}
   \label{fig1}
\end{figure}

The sensitivity of EXPLORER and NAUTILUS during 1998 was not a very good one, worse than that 
obtained in the following years. The pulse sensitivity for 1 ms bursts is of the order of 
$h\,\sim\,1.5\,\times\,10^{-18}$ for EXPLORER and  of $h\,\sim\,10^{-18}$ for NAUTILUS.  
In Fig. \ref{fig1} we show for the two detectors the distribution of the $T_{\rm eff}$ values 
associated to each event, obtained by averaging the filtered data during the ten minutes preceding 
each event.

\section{Search for coincidences}
In the previously published paper (Astone et al. 2001) we found a small coincidence excess 
during  1998 ($n_c = 61$, $\bar{n} = 50.5$) when the detectors were favourably oriented towards 
the Galactic Centre. The coincidence search was based on the use of an energy filter consisting 
in verifying that the two measured energies of the coincidence events be both compatible within 
68\% with the same excitation (see Astone et al. 2001 for details). 

\begin{figure}
   \resizebox{\hsize}{!}{\includegraphics{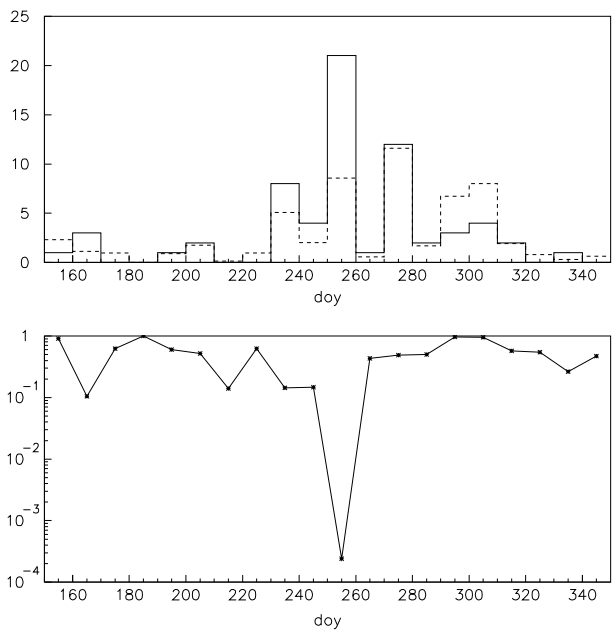}}
   \caption{In the upper graph we show the number of coincidences 
            $n_c$ (full line) and the average number of accidentals 
		$\bar{n}$ (dashed line) for the various ten-day periods. 
		In the lower graph we show the Poisson probability that, 
		given the average $\bar{n}$, we have by chance a number 
		of coincidences equal or grater than $n_c$.}
   \label{fig2}
\end{figure}

With the present paper we have decided to study in more detail this small coincidence excess, by 
dividing the entire period of analysis from 2 June 1998 through 13 December 1998 in ten-day 
periods and applying to each period the same coincidence search as in Astone et al. 2001 with the 
same coincidence window $w = \pm$ 1 s and considering all events with $T_{\rm eff} \le $ 100 mK. 
The result is shown in Fig. \ref{fig2}.
We notice that a large fraction of the small coincidence excess already found is concentrated 
in the period 250-260 day (7-17 September 1998), where we found $n_c = 21$, $\bar{n} = 8.14$.

\begin{figure}
   \resizebox{\hsize}{!}{\includegraphics{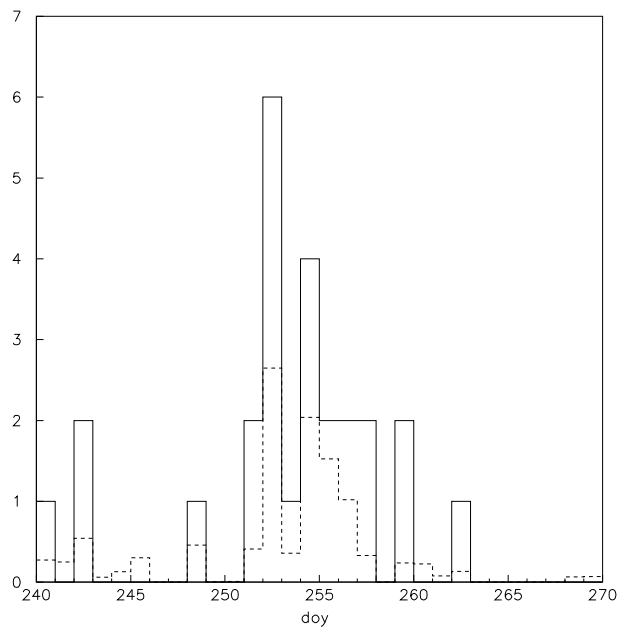}}
   \caption{Day-by-day average number $\bar{n}$ of accidentals (dashed 
            line) and number $n_c$ of coincidences (solid line) in the  
		period 28 August-27 September 1998.}
   \label{fig3}
\end{figure}

If we  investigate in more detail the coincidences in a period including the ten days, that is 
day by day, we find the result given in the Fig. \ref{fig3}.

\begin{figure}
   \resizebox{\hsize}{!}{\includegraphics{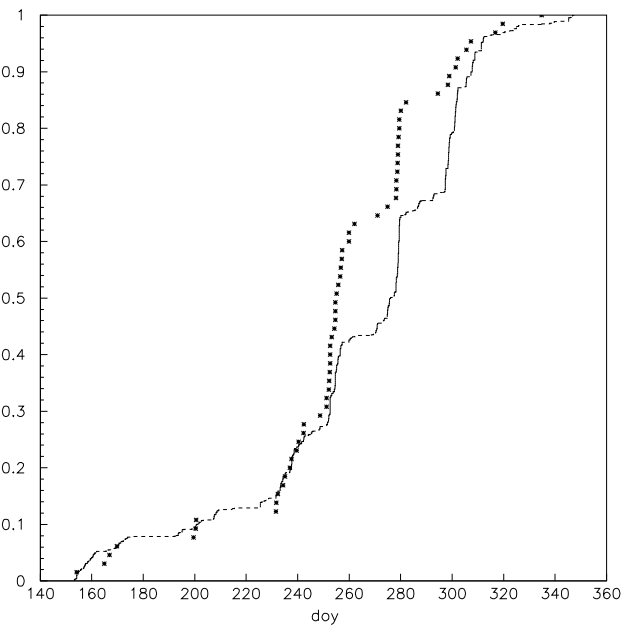}}
   \caption{Cumulative distribution for the coincidences (asterisks) and 
            the accidentals (dashed line) versus the day of the year.}
   \label{fig4}
\end{figure}

We have evaluated, by means of the Kolmogoroff test, the probability that the  distribution 
of the coincidences, as shown in Fig. \ref{fig2}, be a background fluctuation. The cumulative 
distribution is given in Fig. \ref{fig4}, showing that the probability the coincidence 
distribution being a background fluctuation is $P_{\rm Kolm} = 0.5$ \%.

\begin{figure}
   \resizebox{\hsize}{!}{\includegraphics{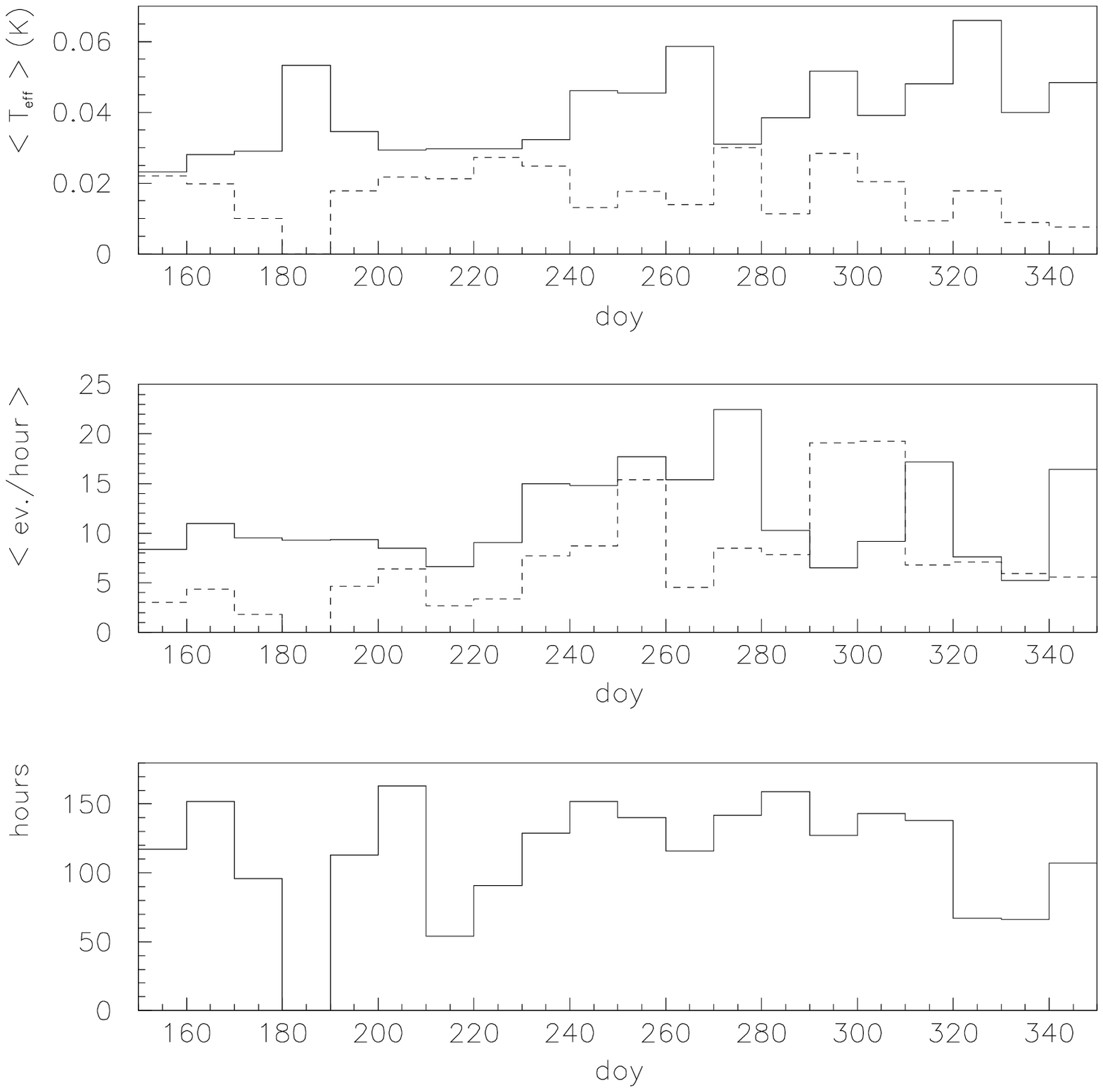}}
   \caption{In the upper graph we show $T_{\rm eff}$ averaged over each 
            ten-day period for EXPLORER (full line) and NAUTILUS (dashed 
		line). In the middle graph the average number of events per 
		hour is given for the various ten-day periods. In the lower 
		graph the number of common hours of operation for each ten-day 
		period is given.}
    \label{fig5}
\end{figure}

It is important to verify whether the operational conditions of the apparatuses in that period 
were not such as to cause this abnormal behaviour. We show in Fig. \ref{fig5} the history of 
$T_{\rm eff}$ averaged over each ten-day period for EXPLORER and NAUTILUS, the average number 
of events per hour and the number of common hours of operation  for the various ten-day period. 
By inspecting this figure we notice a coverage of about 50\% for most ten-day periods and a 
varying number of events per hour, but not such to justify any special behaviour in the period 
250-260 days.

\begin{figure}
   \resizebox{\hsize}{!}{\includegraphics{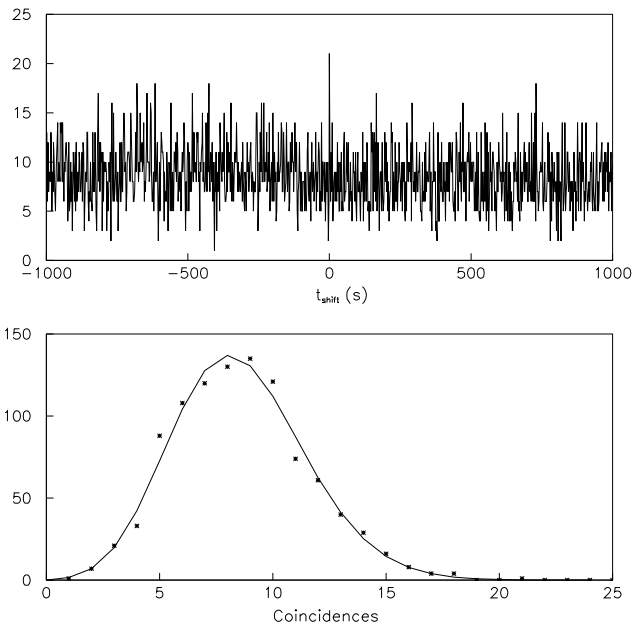}}
   \caption{Delay histogram (upper graph) and Poisson distribution 
            (lower graph) of one thousand delayed coincidences for the 
		period 7-17 September 1998. We have $n_c = 21$ coincidences 
		at $zero$ delay, and an average number of accidental 
		coincidences $\bar{n} = 8.14$.}
   \label{fig6}
\end{figure}

A very important test for verifying that the observed number of accidental coincidences is not, 
with high probability, a background fluctuation and that the apparatus is properly working can 
be done by studying the delay histogram with a reasonably sufficient number of delays for the 
estimation of the accidentals (Astone et al. 2000). The  accidentals have been obtained by time 
shifting one of the two event list with respect to the other one by time steps of 2 seconds 
from -1000 s to +1000 s,  as shown in Fig. \ref{fig6}. We notice that the accidental behaviour 
is good,  indicating that the coincidence excess at time $zero$ is statistically significant. 
The statistical distribution of the accidental coincidences is shown in the lower part of the 
Fig. \ref{fig4}. The asterisks  indicate the measured occurrence of accidentals, the continuos 
line the Poissonian distribution obtained theoretically from the determined average number 
$\bar{n} = 8.14$ of accidentals: the agreement is good.

In our previous paper (Astone et al. 2001) we analysed the data taking into consideration that, 
as the Earth rotates around its axis during the day, the detector happens to be variably oriented 
with respect to a given source. Thus we expect the signal to be modulated during the day; more 
precisely the modulation is expected to have a period of one sidereal day, since the GW sources, 
if any, are  certainly located far outside our Solar system.

\begin{figure}
   \resizebox{\hsize}{!}{\includegraphics{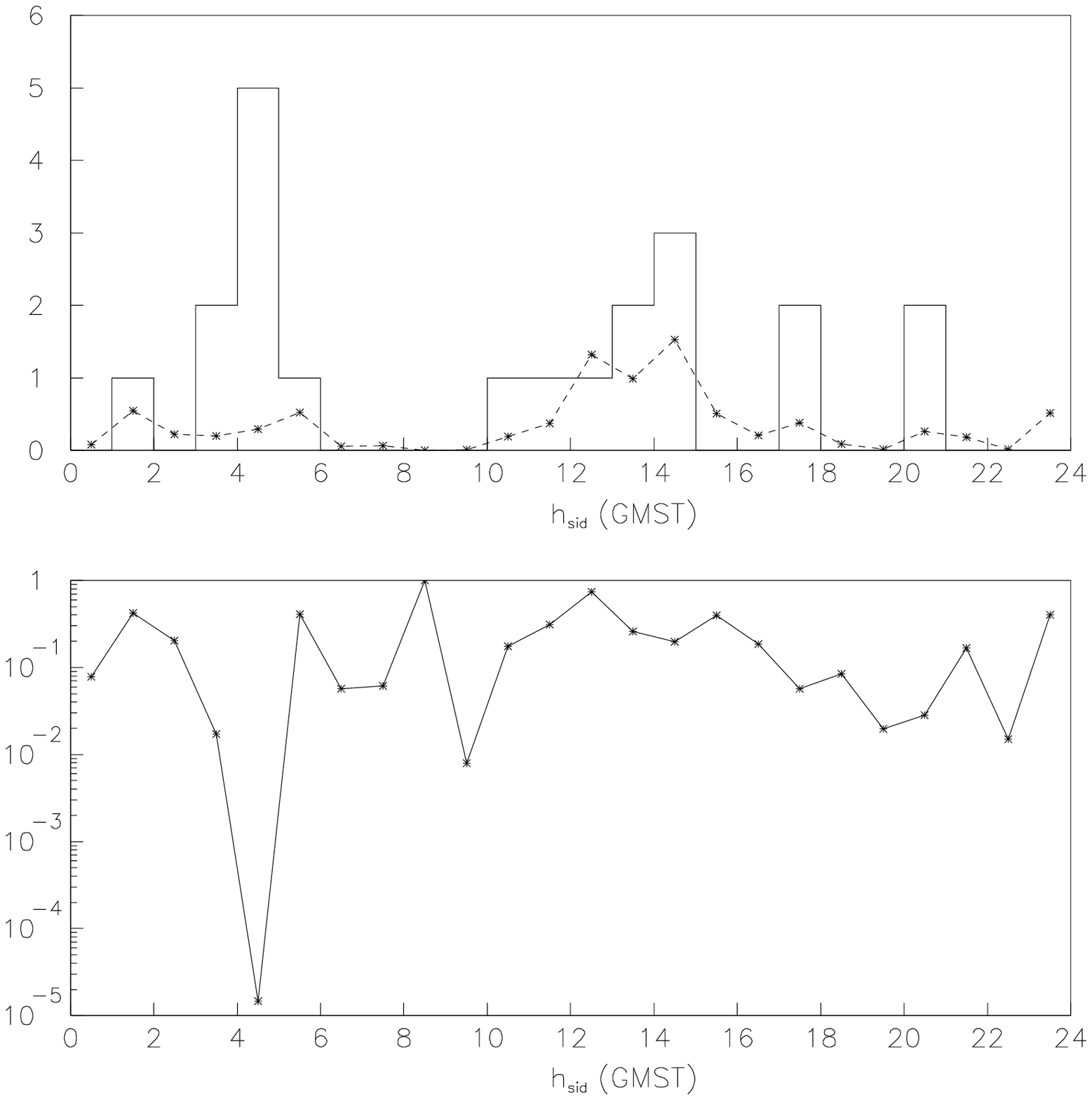}}
   \caption{In the upper graph we show the coincidences $n_c$ (full 
            line) and the average accidentals $\bar{n}$ (dashed line) 
	      versus the sidereal hour for the period 7-17 September 1998. 
		In the lower graph we show the Poissonian probability that, 
		given the average $\bar{n}$ we have by chance a number of 
		coincidences equal or greater than $n_c$.}
   \label{fig7}
\end{figure}

We proceed here just as done for the 2001 data. Thus we search for coincidences at each sidereal 
hour. For the calculation of the sidereal hour we use the Greenwich time, instead of the 
EXPLORER-NAUTILUS local time (longitude = 9.46$^{\rm o}$), as done in the paper Astone et al. 2001.  
The time difference is of about 38 minutes. The 1998 result is shown in Fig. \ref{fig7}. 

We notice that the coincidence excess occurs at the same sidereal time as found with the 2001 
data, when the detectors are well oriented with respect to the Galactic Disc. It turns out that 
at this time of the year sidereal and solar hour almost coincide: at solar hour 3.5 we have 
sidereal hour 4. The peak shown in Fig. \ref{fig7}  has a well defined physical meaning with 
respect to the Galactic Disc, as already found with the 2001 result. 

None of the detected coincidences happens to be at a time when the cosmic ray detector operating 
on NAUTILUS indicates the arrival of a cosmic ray shower.

\section{Conclusion}
The small coincidence excess between the GW detectors EXPLORER and NAUTILUS, already found 
(Astone et al. 2001) with the 1998 data, is  concentrated in the ten-day period 7-17 September 
1998, when the coincidence excess becomes remarkably large. We have checked that  the operational 
conditions of the apparatuses were not such as to justify a special behaviour in those days 
and, particularly, we checked that the distribution of the delayed coincidences is well behaved, 
supporting the statistical significancy of the coincidences excess. 

Nevertheless a warning must be made on the probability estimation. The ten-day period 7-17 September 
has been chosen \emph{a posteriori}. Thus any probability figure has to be taken with care. However, 
we remark that the distribution of Fig. \ref{fig2} has small probability to be a background 
fluctuaction and that the coincidence excess occurs at the same sidereal hours found in 2001.

A problem is encountered if the signal amplitude is considered. In 1998 the signals had energies 
of the order of one half kelvin or more ($h \ge 10^{-17}$), larger than those found with the 
2001 data analysis (Astone et al. 2002). However the signals in 1998 are concentrated in a short 
time interval with a special sidereal time signature, suggesting that the phenomenon, if any, is 
local both in space and time, and therefore it may not be expected to happen again in a few year 
time scale.

\end{document}